# Comparing cars with apples? Identifying the appropriate benchmark countries for relative ecological pollution rankings and international learning

Dominik Hartmann[1,2,3,*], Diogo Ferraz[2,4,5], Mayra Bezerra[3], Andreas Pyka[2], and Flávio L. Pinheiro[6]

[1] Department of Economics and International Relations, Federal University of Santa Catarina R. Eng. Agronômico Andrei Cristian Ferreira, s/n - Trindade, Florianópolis, SC, 88040-900, Brazil
[2] Innovation Economics, Institute of Economics, University of Hohenheim, Germany
[3] Fraunhofer Center for International Management and Knowledge Economy, Neumarkt 9, 04109 Leipzig, Germany
[4] Department of Economics, Federal University of Ouro Preto (UFOP), Mariana 35420-000, Brazil
[5] Department of Production Engineering, São Paulo State University (UNESP), Bauru, Brazil
[6] Nova Information Management School (NOVA IMS), Universidade Nova de Lisboa, PT Campus de Campolide, 1070-312 Lisboa, Portugal
* Corresponding author: dominik.hartmann@ufsc.br

## Abstract

Research in Data Envelopment Analysis has created rankings of the ecological efficiency of countries' economies. At the same time, research in economic complexity has provided new methods to depict productive structures and has analyzed how economic diversification and sophistication affects environmental pollution indicators. However, no research so far has compared the ecological efficiency of countries with similar productive structures and levels of economic complexity, combining the strengths of both approaches. In this article, we use data on 774 different types of exports, $CO_2$ emissions, and the ecological footprint of 99 countries to create a relative ecological pollution ranking (REPR). Moreover, we use methods from network science to reveal a benchmark network of the best learning partners based on country pairs with a large extent of export similarity, yet significant differences in pollution values. This is important because it helps to reveal adequate benchmark countries for efficiency improvements and cleaner production, considering that countries may specialize in substantially different types of economic activities. Finally, the article (i) illustrates large efficiency improvements within current global output levels, (ii) helps to identify countries that can best learn from each other, and (iii) improves the information base in international negotiations for the sake of a clean global production system.

# 1. Introduction

Countries are facing challenges in promoting economic growth without negatively impacting the environment. Due to the threat of climate change, increasing levels of global pollution, deteriorating of natural habitats and biodiversity, and their negative effects on economies and human societies, governments, and international agencies are increasingly aiming to reduce pollutant emissions and the use of resources. For instance, the United Nations created 17 Sustainable Development Goals (SGDs) as a guide to achieve sustainable development outcomes (Griggs, 2013; Robert et al., 2005); the Paris Agreement aimed at raising awareness of worldwide climate change ( United Nations and Framework Convention on Climate Change, 2015); and the World Economic Forum has recently highlighted sustainable development as a key global challenge (World Economic Forum, 2020; 2021). In consequence, several different sustainability indicators and rankings, such as the ecological footprint (Costanza, 2000), greenhouse gas emissions (Hammitt et al., 1996), and ecological efficiency (Camarero et al., 2013), have been created.

In international climate and sustainability summits and negotiations, governments often emphasize different aspects of environmental damage—such as cumulative pollution values, absolute pollution values, or production efficiency—and point to respective indicators. At the same time, differences in economic development levels are tension points between developing, emerging, and mature industrialized countries. Developing countries frequently challenge developed nations to reduce their absolute levels of greenhouse gas emissions and point to their need for economic catch-up and industrialization. In contrast, developed countries often argue that developing regions must promote cleaner technologies and ecological efficiency from the outset of economic development.

Surprisingly, few sustainability rankings take into consideration differences in productive structures between countries beyond similarities in aggregate Gross Domestic Product (GDP). Indeed, specialized literature has either focused on environmental damage effects stemming from economic growth, urbanization, energy resources, and economic sophistication (Dinda, 2005; Sharma, 2011; Martínez-Zarzoso and Maruotti, 2011; Bakhsh et al., 2017; Shahzad et al., 2020), or on the role of different types of pollutants and ecological efficiency based on aggregate economic growth (Camarero et al., 2013; Camioto et al., 2014; Vencheh et al., 2005). Less attention has been given to the potential of cooperation between countries with similar productive structures and on the identification of appropriate benchmark countries for international learning and knowledge transfer. Similarities in productive structures and sophistication are important, though, because as countries specialize in different types of agriculture, industry, or services, they require, by definition, different types of resources linked with different types of pollutants. Thus, countries may not only benefit from learning from the most technologically advanced countries, but also from countries facing similar current productive challenges and opportunities.

Research in economic complexity has shown that countries—and especially developing and emerging economies—do not diversify randomly into new activities, but rather are strongly constrained by their existing productive specialization structure (Hidalgo et al., 2007, Pinheiro et al., 2018; Hartmann et al., 2020, 2021). This implies that a country specialized, for instance, in agricultural products, textile products, or highly-polluting and energy-intensive products, such as steel or aluminum, may not easily transition into producing wind or solar energy technologies as a new base of their economy. However, it also means that comparing pollution values of countries with very different productive structures may not be the best comparative benchmark to understand which countries show a relatively clean or polluting production system and which countries could best learn from one another. Simple comparisons based on absolute pollution indices or GDP might end up comparing “apples and cars”, instead of like with like. For instance, the USA has significantly higher pollution values than both Japan and Madagascar, but its export portfolio is much more similar to Japan than to Madagascar (see Figure 1). Japan ($CO_2$ emissions per capita = 9.54 metric tons per capita) is arguably a better benchmark country for the USA (16.50) to learn about ecologically more efficient and cleaner technologies for its product portfolio than Madagascar (0.13). Moreover, while Japan produces significantly lower levels of pollution for a similar export portfolio and level of economic complexity than the USA, thus has an ecologically more efficient production system, the same cannot be claimed in a straightforward manner for the comparison between Japan and Madagascar. Thus, appropriate benchmark countries need to be identified to evaluate the eco-efficiency of countries and to identify best learning partners.

To reveal adequate benchmark countries—in our case, countries that export similar products but have significantly different pollution values—we combine methods from network science, economic complexity research, and data envelopment analysis (DEA). Such a multimodal approach helps to create sustainability rankings that take productive specialization of countries into account and identify potentials for sustainability improvements within the current global export and production systems. It is important to note that this article does not focus on green diversification opportunities, which is an important topic scrutinized elsewhere (Fraccascia et al., 2018; Dordmong et al., 2021), but focuses on the current relative benchmarks and potentials for efficiency improvements and learning.

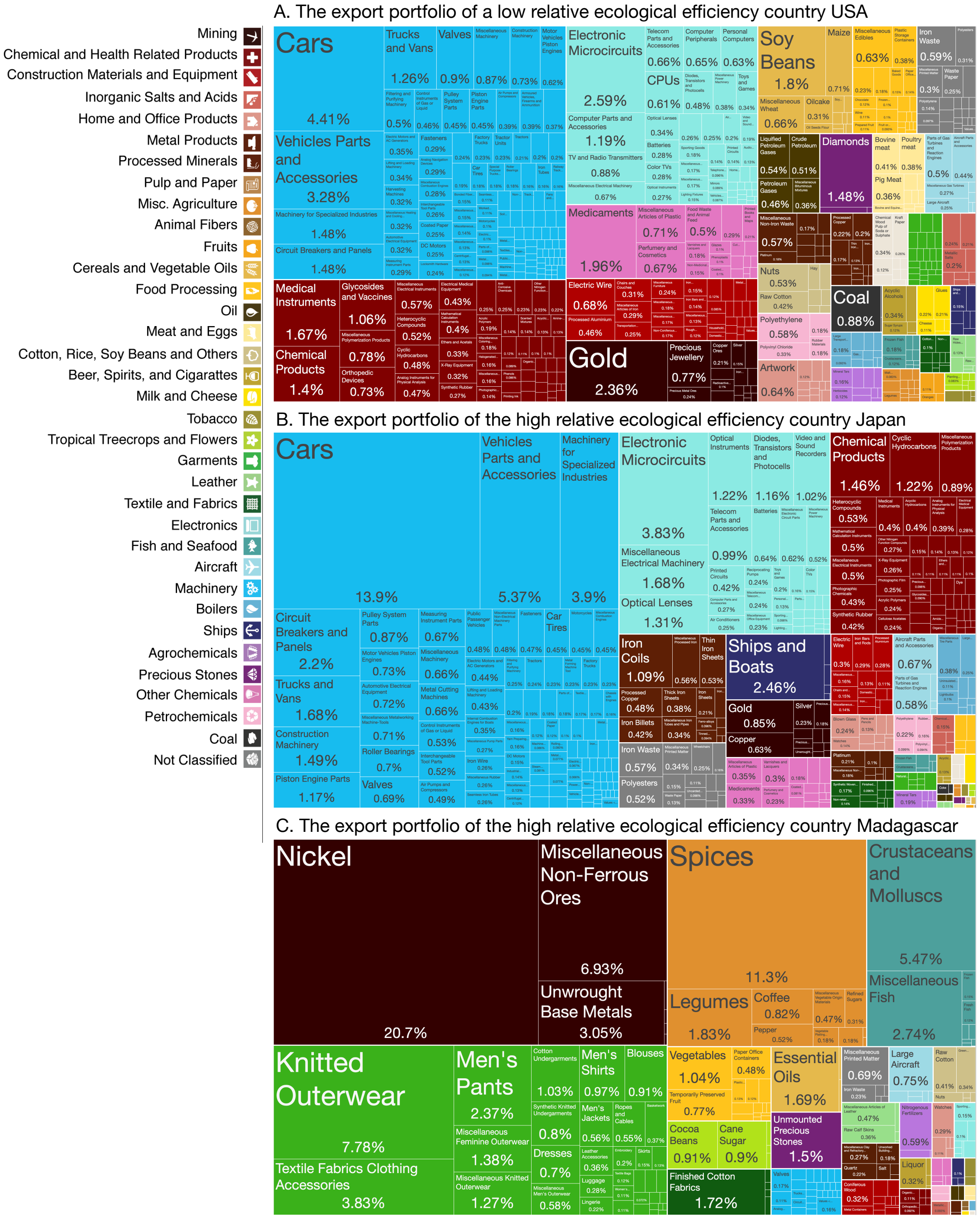

**Figure 1.** Example of the export portfolios of the USA (top), Japan (middle), and Madagascar (bottom). Products are colored according to their category class, and the area is proportional to the share of exports for each country. Source: oec.world, own illustration.

The remainder of the article proceeds as follows. First, we review the literature on economic complexity and sustainability indicators. Then we present our data and methods. In the results section, we first discuss the economic development weighted sustainability ranking and then present a network that shows the best sustainability benchmark countries. Finally, we quantify the overall efficiency improvement potential if all countries would move to the efficiency frontier. Naturally, the study has its limitations, such as exports being a proxy indicator for productive specialization, or the fact that not all countries are necessarily able to produce the same products with the same combination of inputs. However, we argue, it is a valid step forward to consider significant differences in productive structures when comparing their sustainability levels and identifying promising countries that could learn from one another for the sake of more efficient and cleaner production systems.

# 2. Literature Review

Several studies argue that there is a direct link between economic growth and the emission of pollutant gases (Chan and Yao, 2008; Fujii and Managi, 2016; Zhang and Cheng, 2009; Li et al., 2014). This perspective emphasizes that alternative growth strategies are required to increase GDP with less pollution (Hashmi and Alam, 2019). In contrast, the Environmental Kuznets Curve (EKC) hypothesis proposes an inverted U-shape between economic growth and emissions that implies a reduction of environmental impacts at higher levels of GDP (Dinda, 2005). Other studies argue that urbanization affects environmental degradation in several ways, and indicate that urbanization increases pollutant emissions, while others also show that urbanization might contribute to environmental improvements (Poumanyvong and Kaneko, 2010; Sharma, 2011). Finally, some studies focus on the importance of alternative energy resources, such as wind turbines and photovoltaic cells, to reduce environmental degradation (Pegels and Lütkenhorst, 2014; Scarlat et al., 2015). Indeed, clean energy might reduce the use of fossil fuels, resulting in a reduced impact on the environment. Nordic countries and other

European regions are examples of a successful energy transformation (Bakhsh et al., 2017; Huynh and Hoang, 2019). However, countries with no access to these technologies face difficulties using clean energy, which shows the importance of considering the technology and productive structure of economies, especially in a development context.

In this regard, research on economic complexity shows that countries with a diversified and complex productive structure can use technology to reduce the ecological damage (Doğan et al., 2019; Shahzad et al., 2020). The Economic Complexity Index (ECI) evaluates the diversification and sophistication of the productive structure (Hidalgo, 2021; Hidalgo and Hausmann, 2009; Hidalgo et al., 2007).

Past works has illustrated that countries with a high level of economic complexity have the necessary capabilities to produce green products, such as electric cars, clean energy, among others (Casals et al., 2016; Gangale et al., 2017; Fraccascia et al., 2018). These studies analyze different countries (e.g. France and Turkey) and regions (e.g. Europe), and divided countries according to income or development groups (Shahzad et al., 2020). Their main finding is that economic complexity is an alternative way to promote economic growth while reducing pollutant emissions (Can and Gozgor, 2017; Doğan et al., 2019; Gozgor and Can, 2016; Neagu, 2019, 2020; Neagu and Teodoru, 2019; Shahzad et al., 2020). Conversely, it is important to note that there is also evidence that points to a potential increase in the emissions of particular types of pollutants with increasing levels of complexity (Boleti et al., 2021). Additionally, the potential outsourcing of more polluting economic activities may not necessarily reduce the environmental damage caused by the world´s production system, if it does not increase the overall ecological efficiency of the production of these goods or services. Finally, despite analyzing the nexus between economic sophistication and environmental damage (Ferraz et al., 2021), the literature tends to neglect sustainability indicators, especially those referring to ecological efficiency.

Eco-efficiency indicators reveal countries that promote economic growth with less environmental degradation. This is important because it helps to understand the best practices and to identify the right benchmarks countries, especially for countries with high levels of environmental damage. Techniques from Data Envelopment Analysis (DEA) support this strand of research to compare eco-efficiency ranking positions of countries and regions by using different proxies for environmental degradation, such as carbon dioxide ($CO_2$), nitrogen oxides (NOX), and sulfur oxides (SOX) (Camarero et al., 2013), specific regions (i.e. Latin America) (Moutinho et al., 2018), and economic sectors (Camioto et al., 2014; Zhang et al., 2008). The main finding of the eco-efficiency literature indicates that only a limited number of countries (i.e., Switzerland and Scandinavian countries) can be considered eco-efficient. In contrast, several parts of the globe, such as (south-)eastern European countries (e.g. Hungary and Turkey), North-America (Canada and the United States) and Latin America (Moutinho et al., 2018) are characterized by low levels of efficiency (Camarero et al., 2013). These findings show that eco-inefficient countries face severe difficulties in developing cleaner production. Arguably cooperation and knowledge transfer with eco-efficient countries could help in this regard.

So far, the eco-efficiency studies have put less emphasis on the productive structures and sophistication of economic systems, which, however, are relevant to compare countries and identify countries that can learn best from each other. Most studies use GDP as the indicator for the economic development of countries or regions. However, the aggregated GDP measure potentially hides substantial differences in particular technological and ecological challenges. For example, country A specialized in agricultural products, country B specialized in crude petroleum, and country C specialized in textile industries; while they could have similar levels of GDP, they might not be the best countries to learn from each other to improve the ecological efficiency. While DEA indicators allow for a comparison of the pollution efficiency of countries with similar levels of GDP and average economic complexity, they are not sufficient to identify which countries could learn from each other and thereby could make their production systems more sustainable. To that end, the type and composition of products that countries are producing also need to be taken into consideration.

Economic structures embody the knowledge that exists in production systems and condition a country's level of pollution as well as its green growth opportunities (Fraccascia et al., 2018; Hidalgo et al., 2007). Similarities in productive structures between countries are crucial for effective knowledge transfer and are essential in predicting their absorptive capacities to learn from each other (Cummings and Teng, 2003). Knowledge transfer builds on the ability of economic agents and international organizations to transfer innovation and technology to other countries in meaningful ways (Cummings and Teng, 2003). Knowledge transfer is a basis for comparative advantages over the years (Argote and Ingram, 2000) and depends on the period of partnerships (Håkanson and Nobel, 2000; 2001), as well as the available budget and the structure of the production systems (Pinto and Mantel, 1990; Szulanski, 1996). Moreover, research from the economic catching-up literature showed that the bell-shaped relation between the technological gap and the ability to transfer external knowledge can explain the large possibilities of lagging-behind countries to learn from leading countries (Verspagen, 1992).

Despite the burgeoning literature on economic complexity, environmental degradation, and eco-efficiency, the concepts have not yet been discussed in an integrated manner to improve our understanding of a better reduction of environmental damage. In other words, studies like the article at hand are missing so far, which compare countries' productive structure and environmental degradation with the aim of proposing most meaningful comparative economies to learn from one another.

# 3. Data and methods

We use data on productive structures and environmental damage of 99 developed and developing countries in 2014 to reveal their eco-efficiency and to identify appropriate benchmark countries. Moreover, we use trade data of 774 export goods of the Standard Industrial Trade Classification (SITC) from the Observatory of Economic Complexity (Simoes and Hidalgo, 2011) as proxies for the level of economic sophistication as well as the heterogeneity of the national productive structures (Hidalgo, 2021). In particular, we use exports data to estimate the Economic Complexity Index (ECI) (Hidalgo and Hausmann, 2009) for 110 economies in the year 2014. The ECI measures the knowledge intensity of countries by considering the knowledge intensity embedded in the exported products (Hidalgo and Hausmann, 2009). Due to differences in data availability for countries the trade dataset and the environmental damage dataset, we will focus our analysis on 94 countries (Appendix A shows the analyzed countries).

We start by identifying the differences in the ecological efficiency of countries' economic output. For this purpose, we measure the pollution levels generated by countries to reach certain levels of economic development (proxied by the ECI). Applying methods from Data Envelopment Analysis makes the results comparable with previous estimates from the ecological efficiency literature.

Next, we use a network analysis approach to identify pairs of countries with similar productive structures (proxied by the similarity in achieving revealed comparative advantages of 432 non-primary goods exports) and substantial differences in ecological efficiency. The network approach presents two advantages. The partnership-network method allows for the analysis of hundreds of economic sectors present in the 94 analyzed countries and identification of the best benchmark countries. Moreover, the obtained partnership network provides a better framework to visualize the interplay between export similarity with the potential for mutual learning and efficiency improvements through learning and knowledge transfer between countries. In other words, the resulting partnership network is considered as the benchmark to identify the best country learning partnerships allowing for sustainability improvements.

We use two variables to represent environmental degradation: $CO_2$ emissions and ecological footprint. Carbon dioxide ($CO_2$) emissions, measured in metric tons per capita (World Bank, 2019a), stem from burning of fossil fuels and manufacturing of cement, and include carbon dioxide produced during the consumption of solid, liquid, and gas fuels and gas flaring. The ecological footprint (EF) measures how much biologically productive land and water an individual, a population, or an activity requires to produce all the resources it consumes and also absorbs the waste it generates, using prevailing technology and resource management practices (Costanza, 2000; Wackernagel and Rees, 2004; Fiala, 2008). Several studies have argued that the ecological footprint is an important global and comparable indicator for environmental degradation, which is affected by income, trade openness, energy, and renewable resources (Charfeddine, 2017; Destek et al., 2018; Zafar et al., 2019; Neagu, 2020). We use the Ecological Footprint of consumption in global hectares (gha) divided by population (EFConsPerCap). Accordingly, we analyze how countries generate environmental degradation, taking also their underlying productive structure into account.

To estimate the eco-efficiency of countries considering their level of economic sophistication, we use methods from Data Envelopment Analysis. It must be noted that most studies on ecological indicators using DEA are concerned with technical issues, such as weight restrictions, model orientation, and desirable and undesirable outputs. Lovell et al. (1995) present an extended additive model to interpret better relative efficiency. For this, the authors transform undesirable outputs (i.e. $CO_2$ emissions) using a translation technique by adding a large scalar to the additive inverse (i.e., multiplication by −1). This translation approach is necessary because it allows positive values for each analyzed unit (Lovell et al., 1995). Färe et al. (1996) measure environmental performance by using the ratio between the reduced undesirable output and the increased quantities of inputs or the decreased quantities of desirable outputs (Färe et al., 1996). Other studies treat the undesirable pollutant emissions output as a classical DEA input (Camioto et al., 2016; Camioto et al., 2014; Korhonen and Luptacik, 2004). Vencheh et al. (2005) develop a DEA model to treat undesirable inputs and outputs simultaneously (Vencheh et al., 2005). It is important to note that these studies have not yet considered new variables on the eco-efficiency analysis, such as the ecological footprint and countries' productive structures and levels of economic complexity.

Regarding the DEA approach, we use a Variable Return of Scale (VRS) model. The environmental degradation variables are considered undesirable outputs, which must be treated before achieving ecological efficiency. We follow several studies that treat pollutant emissions as a classical DEA input (Camioto et al., 2016; Camioto et al., 2014; Korhonen and Luptacik, 2004). This approach presents a more intuitive analysis, since the original data is used and the minimization of environmental degradation is directly considered (Dyckhoff and Allen, 2001). In other words, our DEA model was programmed to decrease pollution and ecological footprint inputs maintaining the same level of economic sophistication (Korhonen and Luptacik, 2004; Kuosmanen and Kortelainen, 2005).

The resulting Relative Ecological Pollution Ranking (REPR) measures eco-efficiency considering the level of economic complexity of the countries. Our REPR shows the efficiency of countries in achieving high levels of economic complexity based on relatively low levels of $CO_2$ emissions and ecological footprints. The REPR is calculated as follows:

**Table 1** – Data Envelopment Analysis (DEA) radial model in the form of multipliers

**Relative Ecological Pollution Ranking (REPR)**

$$\max \sum_{i=1}^{m} u_i . y_{i0} + w$$

subject to:

$$\sum_{j=1}^{n} v_j . x_{j0} = 1$$

$$\sum_{i=1}^{m} u_i . y_{ik} - \sum_{j=1}^{n} v_j . x_{jk} + w \leq 0 \; for \; k = 1,2,\dots,h$$

$w$ without sign restriction.

**Source**: Mariano and Rebelatto (2014, p. 5)

Where: $x_{jk}$ represents the amount of the environmental variables ($CO_2$ emissions and ecological footprint) j of a country $k$; $y_{ik}$ represents the amount of the economic complexity $i$ of a country $k$; $x_{j0}$ represents the amount of the environmental variables j of the country; $y_{i0}$ represents the amount of economic complexity I of the country; $v_j$ represents the weight of the environmental variables $j$ for the country; $u_i$ represents the weight of the economic complexity $i$ for the country; $\theta$ means the efficiency of the country being analyzed; $\lambda_k$ is the contribution of the country $k$ to the goal of the region; $m$ is the quantity of analyzed economic complexity; $n$ is the quantity of environmental variables analyzed; and $W$ represents the scale factor. In this sense, countries with a value equal to zero have the lowest relative ecological pollution performance, while countries with a value equal to one have the highest relative ecological pollution performance.

### 3.1 Export similarity and potential improvement of production efficiency

In the next step, we calculate the network of similarities between countries' export baskets in order to identify partners with substantial differences in the eco-efficiency while achieving similar levels of economic sophistication. To that end, we compare the logarithm of the revealed comparative advantage (RCA) of countries' product basket. We start by computing the RCA of each country on each product as:

$$R_{cp} = \frac{X_{cp}}{\sum_{p'} X_{cp'}} \Big/ \frac{\sum_{c'} X_{c'p}}{\sum_{c'p'} X_{c'p'}} \tag{3}$$

where $X_{cp}$ is the total exports of country $c$ over product $p$. The ratio in the numerator estimates the relative weight of exports of a product $p$ in the economy of country $c$, while the ratio in the nominator estimates the relative weight of product $p$ in the world economy and thus represents the weight of product $p$ in a typical/ average country. By definition, RCA is bounded within the domain of positive real numbers that are right-skewed distributed. In order to obtain linearly comparable *country-to-country* RCAs, we apply the log-transform to $R_{cp}$ as:

$$\tilde{R}_{cp} = \log_{10}(R_{cp} + \delta) \tag{4}$$

where the sum of $\delta$ is calculated to ensure that undefined transformations are avoided from instances where $R_{cp} = 0$; in our case we considered $\delta$ to be the smallest non-zero value of $R_{cp}$. Hence, $\tilde{R}_{cp}$ quantifies the magnitude of revealed comparative advantages. Figure 2a) shows the distribution of values of $\tilde{R}_{cp}$ obtained for all countries.

Hence, for each country $c$ we obtain a vector $\tilde{\boldsymbol{R}}_{\boldsymbol{c}}$: $\{\tilde{R}_{c1}, \tilde{R}_{c2}, \ldots, \tilde{R}_{cN}\}$ that captures the magnitude of the revealed comparative advantages per product from country $c$. Next, we compute the correlations between the magnitudes of revealed comparative advantages ($\tilde{\boldsymbol{R}}_{\boldsymbol{c}}$) from each pair of country $c$ and $c'$. For this we compute the Pearson correlation coefficient between vectors $\tilde{\boldsymbol{R}}_{\boldsymbol{c}}$ and $\tilde{\boldsymbol{R}}_{\boldsymbol{c'}}$, which is formally:

$$\rho_{cc'} = \frac{\sum_i^N (\tilde{R}_{ci} - \langle \tilde{\boldsymbol{R}}_{\boldsymbol{c}} \rangle)(\tilde{R}_{c'i} - \langle \tilde{\boldsymbol{R}}_{\boldsymbol{c'}} \rangle)}{\sqrt{\sum_i^N \left(\tilde{R}_{ci} - \langle \tilde{\boldsymbol{R}}_{\boldsymbol{c}} \rangle\right)^2} \sqrt{\sum_i^N \left(\tilde{R}_{c'i} - \langle \tilde{\boldsymbol{R}}_{\boldsymbol{c'}} \rangle\right)^2}} \tag{5}$$

where $\langle \tilde{\boldsymbol{R}}_{\boldsymbol{c}} \rangle$ is the mean magnitude of revealed comparative advantages of country $c$'s exports. Figure 2b) shows the resulting correlation matrix, while Figure 2c) shows an example of the correlation between the exports magnitude of South Korea and Japan. Figure 2d) shows the distribution of correlations.

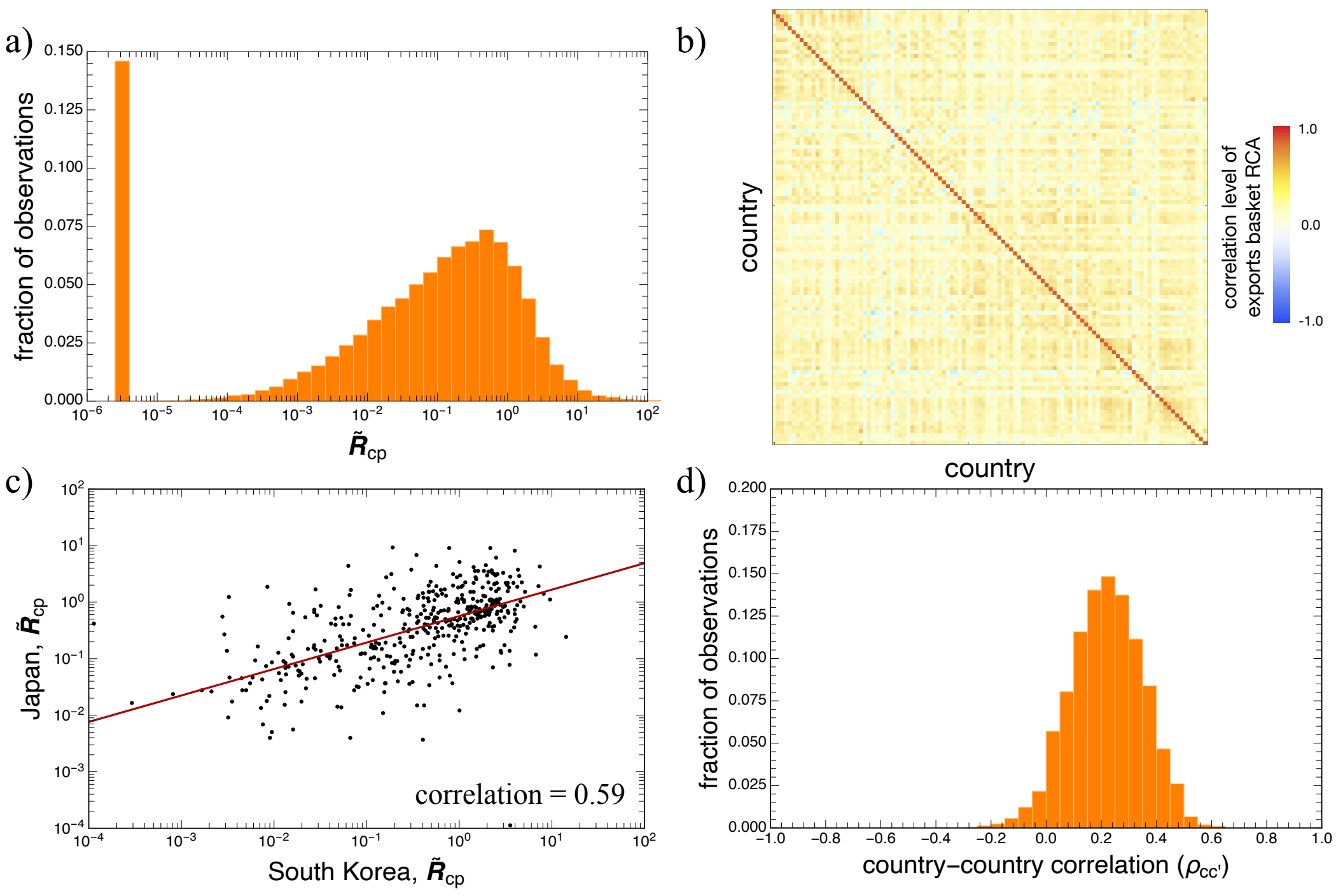

**Figure 2** – Correlations between the export portfolios of countries. Panel a) Distribution of Revealed Comparative Advantages. Panel b) Correlation matrix between countries. Panel c) Example of the correlations between the export basket of South Korea and Japan. Panel d) Distribution of measured correlations between the export portfolios of countries.

To improve the visualization of the strongest linkages that emerge from the correlations of the exports magnitude, we apply the following in order to obtain a meaningful network representation:

1. Starting from the correlation matrix $\rho_{cc'}$, we generate the maximum spanning tree, $S$, that identifies the minimum number of edges necessary to obtain a connected network and that maximizes the sum of the correlations between the edges.
2. $S$ is an undirected network that connects pairs of countries.
3. Then we add to $S$ all of the links associated with countries that exhibit a correlation greater or equal to 0.445. This threshold was selected to obtain a network with an average degree of approximately four links, which results in a graphical representation of the network that balances between interpretability and meaning.

These steps follow the methods used to build a network representation of the Product Space (Hidalgo et al., 2007).

# 4. Results

First, we compare the ecological efficiency of countries, considering their levels of economic development (in terms of economic complexity).

We start with descriptive statistics of the absolute values of input and output variables of the relative ecological pollution ranking (REPR) and compare both absolute and relative environmental damage values of low, middle- and high-income countries (Figure 3). By construction, the average value of Economic Complexity Index (ECI) of the 94 analyzed countries is 0.06. The average of $CO_2$ emissions per capita and ecological footprint per capita are 5.39 and 3.65, respectively. High income countries present higher average ECI values (0.75), $CO_2$ emissions (8.72) as well as ecological footprints (5.45). The upper middle- and lower middle-income countries present lower average ECI values ($-0.63$ and $-0.19$, respectively), $CO_2$ emissions (1.88 and 4.48, respectively), and ecological footprint (1.52 and 4.52, respectively). Moreover, the low-income countries present the lowest levels of economic sophistication (ECI = -1.19 on average), $CO_2$ emissions (0.19) and ecological footprint (0.95). Yet even when taking their economic sophistication (i.e. economic complexity value) into account, low-income countries tend to present less environmental damage in relative terms (REPR = 0.13) than upper middle- (0.65) and lower middle-income (0.52) countries as well as high-income countries (0.56).

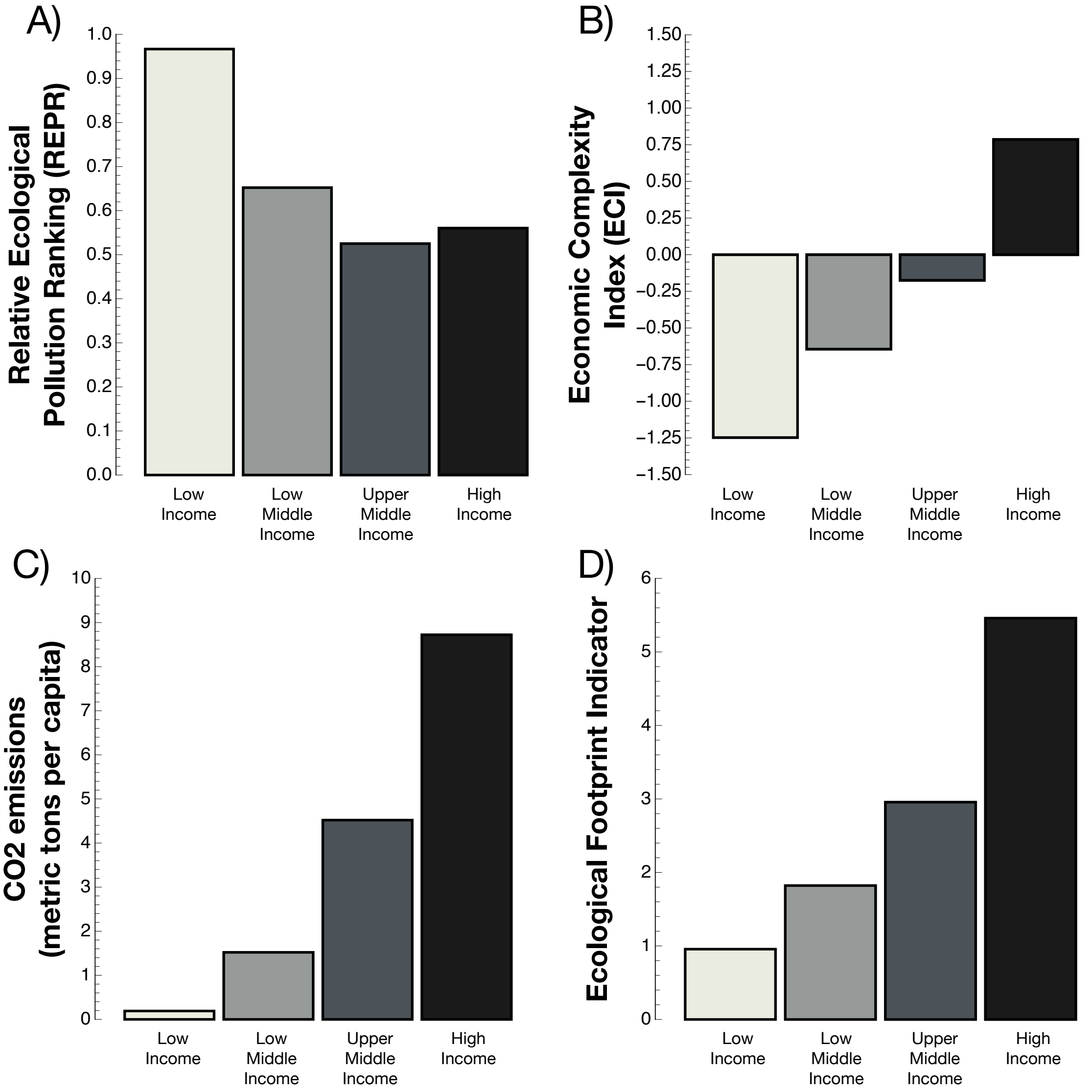

**Figure 3 –** Characterization of the four economic groups according to their relative ecological pollution ranking (REPR), economic complexity, CO2 emissions, and ecological footprint. Income groups follow the convention proposed by the World Bank (2019b).

This means that both in absolute and relative terms, poor countries cause less environmental damage (in terms of their average ecological footprint and $CO_2$ emissions) than rich countries. However, there is also a significant amount of variance, where some rich countries have relatively good relative ecological pollution values and some poor countries pollute relatively more than would be expected from their level of economic sophistication.

Table 2 presents the Top-15 (best performance) and Bottom-15 (worst performance) regions with the relative indicator (the full ranking can be found in Appendix A). The Top-15 countries are mainly composed of low-income and middle lower-income countries. The Top-5 countries are Congo, Dem. Rep. ($1^{st}$), Madagascar ($2^{nd}$), Zambia ($3^{rd}$), Mozambique ($4^{th}$), and Philippines ($5^{th}$). Note that these low-income regions present low levels of economic complexity and relatively less environmental degradation in terms of $CO_2$ emissions and ecological footprint. For example, while the average of the Economic Complexity Index from 99 countries is 0.02, the Top-5 countries present a low sophistication of their productive structure (ECI = -2.83 on average), but the environmental damage is even less on average than in most other countries. For these reasons, these countries have the best performance in the relative indicator. It must also be noted that some high-income countries position among the Top-15 countries with the best performance, such as Switzerland ($7^{th}$), Japan ($9^{th}$), Hungary ($12^{th}$), and Sweden ($15^{th}$). These countries present a high level of economic sophistication (ECI = 1.83), but their average levels of $CO_2$ emissions (5.65) and ecological footprint (4.95) are lower compared to countries with similar levels of economic sophistication. This is quite substantial, especially considering their access to technology.

The Bottom-15 countries presenting the worst relative ecological efficiency values are mostly composed of economies that are closely dependent on natural resources exploitation. The Bottom-5 regions are Kazakhstan (95th), Guinea-Bissau (96th), United Arab Emirates (97th), Mongolia (98th), and Kuwait (99th). These countries present a low level of economic sophistication and substantial environmental damage. For example, Guinea-Bissau has a worse level of economic sophistication and higher levels of $CO_2$ emissions than the average of the low-income country group.

**Table 2** – Top-15 and bottom-15 countries with the best and worst relative ecological indicator (REPR) and their income group in 2014.

| Countries | REPR | Rank | Income Group |
|---|---|---|---|
| **TOP 15 COUNTRIES** | | | |
| Congo, Dem. Rep. | 1.0000 | 1 | Low income |
| Madagascar | 1.0000 | 2 | Low income |
| Zambia | 0.9979 | 3 | Lower middle income |
| Mozambique | 0.9956 | 4 | Low income |
| Philippines | 0.9913 | 5 | Lower middle income |
| Pakistan | 0.9862 | 6 | Lower middle income |
| Switzerland | 0.9567 | 7 | High income |
| Kenya | 0.9489 | 8 | Lower middle income |
| Japan | 0.9391 | 9 | High income |
| Togo | 0.9252 | 10 | Low income |
| Ethiopia | 0.9045 | 11 | Low income |
| Hungary | 0.8873 | 12 | High income |
| Thailand | 0.8384 | 13 | Upper middle income |
| Mexico | 0.8293 | 14 | Upper middle income |
| Sweden | 0.8272 | 15 | High income |
| **BOTTOM 15 COUNTRIES** | | | |
| Estonia | 0.4191 | 85 | High income |
| South Africa | 0.4122 | 86 | Upper middle income |
| New Zealand | 0.4056 | 87 | High income |
| Russian Federation | 0.3682 | 88 | Upper middle income |
| Canada | 0.3407 | 89 | High income |
| Algeria | 0.2786 | 90 | Upper middle income |
| Azerbaijan | 0.2709 | 91 | Upper middle income |
| Saudi Arabia | 0.2311 | 92 | High income |
| Oman | 0.1906 | 93 | High income |
| Australia | 0.1699 | 94 | High income |
| Kazakhstan | 0.1659 | 95 | Upper middle income |
| Guinea-Bissau | 0.1633 | 96 | Low income |
| United Arab Emirates | 0.0029 | 97 | High income |
| Mongolia | 0.0025 | 98 | Lower middle income |
| Kuwait | 0.0000 | 99 | High income |

Next, we use the Country Exports Similarity Space to identify differences in the ecological efficiency of countries (as measured by REPR) with similar export portfolios (see Figure 4) in more detail. To properly learn from another country, it is not enough to consider the average level of sophistication, but also a more fine-grained distinction among types of products. This is the case because countries with a similar level of economic sophistication can base their economy on very different types of productive specialization. One country can focus on chemical products and another on electronic goods, or one country can focus on agriculture and another on mining products. Each of these activities tends to require particular types of productive capabilities and knowledge, but they are also associated with different levels of environmental damage. Figure 4 shows the similarity in the network of countries' export portfolio, with the nodes colored according to their REPR values.

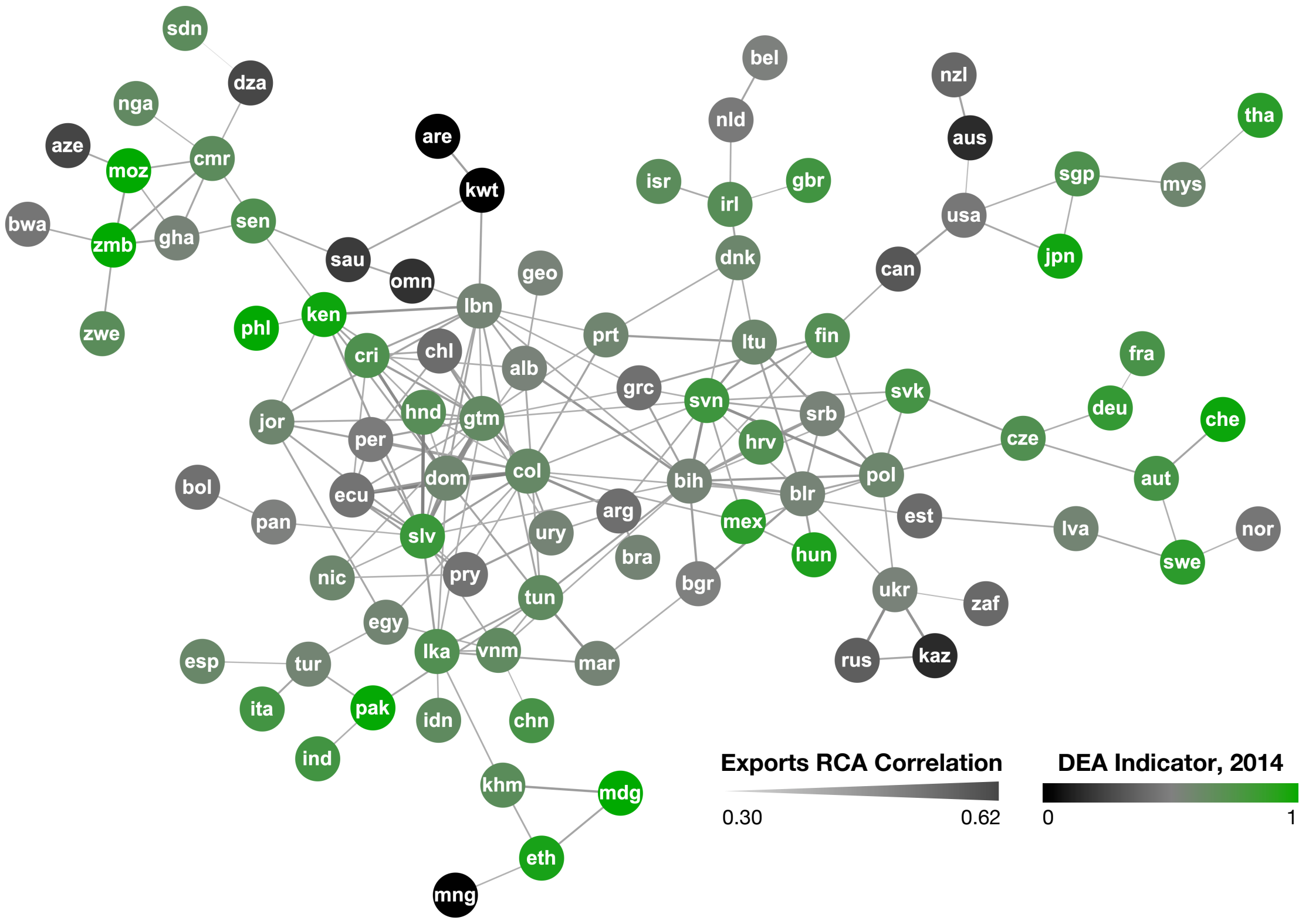

**Figure 4 –** Export similarity network between countries. The nodes are colored according to their relative ecological pollution (REPR) values. The edges are undirected and their thickness and color are scaled to represent exports RCA correlation (thinner and lighter colored means lower correlation, conversely thicker darker edges represent greater correlations). The network was visualized by selecting edges with correlations greater than 0.725 and identifying the edges that form the maximum spanning tree. In doing so, we ensured that the final network has an average degree of approximately four.

Moreover, Table 3 shows the export similarity and REPR values for country pairs with the highest and lowest export similarity. We observe some network clustering of spatial neighbors that share both similarities in terms of export portfolios as well as ecological efficiency, such as France and the United Kingdom, or Saudi Arabia, Kuwait, and Oman. However, there are also considerable differences among neighboring countries, and we can identify major differences in terms of the REPR values of countries with relatively similar export portfolios, such as Japan and the USA, or the Ivory Coast and Cameroon. While these countries are able to export similar type of products and thus reach similar levels of productive sophistication, they show substantial differences in the amount of $CO_2$ emissions and ecological footprint per capita required to reach this level of productive sophistication. This also means that the country with a lower REPR value may be able to learn from the country with a significantly REPR value. They are likely to be a better benchmark country for international comparisons and identification of improvement potentials than studies merely based on aggregate GDP or pollution values across countries with very different productive specializations.

**Table 3** – Country pairs with the highest levels of export similarity. Each row indicates the level of REPR for a focal country ($C_1$) and the partner country ($C_2$) along with the exports correlation ($\rho_{C_1C_2}$) and the differential between the focal and the partner in terms of REPR ($\Delta REPR(C_1, C_2)$).

| Focal Country ($C_1$) | Partner Country ($C_2$) | $\rho_{C_1C_2}$ | $\Delta REPR(C_1, C_2)$ | $REPR_{C_1}$ | $REPR_{C_2}$ |
|---|---|---|---|---|---|
| **Top 20** | | | | | |
| Ecuador | Colombia | 0,62 | 0,15 | 0,45 | 0,60 |
| Honduras | El Salvador | 0,61 | 0,12 | 0,65 | 0,77 |
| Dominican Republic | Costa Rica | 0,58 | 0,13 | 0,56 | 0,69 |
| Dominican Republic | Guatemala | 0,57 | 0,05 | 0,56 | 0,61 |
| Russia | Ukraine | 0,57 | 0,16 | 0,37 | 0,53 |
| Guatemala | El Salvador | 0,56 | 0,16 | 0,61 | 0,77 |
| Bosnia & Herzegovina | Slovenia | 0,56 | 0,22 | 0,54 | 0,75 |
| Colombia | Guatemala | 0,56 | 0,01 | 0,60 | 0,61 |
| Poland | Slovenia | 0,56 | 0,18 | 0,57 | 0,75 |
| Argentina | Colombia | 0,55 | 0,17 | 0,43 | 0,60 |
| Kazakhstan | Ukraine | 0,55 | 0,36 | 0,17 | 0,53 |
| Lebanon | Kenya | 0,54 | 0,41 | 0,54 | 0,95 |
| Albania | Bosnia & Herzegovina | 0,54 | 0,01 | 0,53 | 0,54 |
| Morocco | Tunisia | 0,54 | 0,08 | 0,54 | 0,62 |
| Ecuador | Dominican Republic | 0,53 | 0,11 | 0,45 | 0,56 |
| Chile | Guatemala | 0,52 | 0,19 | 0,42 | 0,61 |
| Poland | Lithuania | 0,52 | 0,00 | 0,57 | 0,58 |
| Bosnia & Herzegovina | Poland | 0,52 | 0,04 | 0,54 | 0,57 |
| Cambodia | Madagascar | 0,52 | 0,35 | 0,65 | 1,00 |
| Lithuania | Slovenia | 0,52 | 0,18 | 0,58 | 0,75 |
| **Bottom 20** | | | | | |
| France | United Kingdom | 0,31 | 0,03 | 0,72 | 0,74 |
| Belgium | Denmark | 0,31 | 0,07 | 0,51 | 0,58 |
| Philippines | Zambia | 0,31 | 0,01 | 0,99 | 1,00 |
| Thailand | Kenya | 0,31 | 0,11 | 0,84 | 0,95 |
| Hungary | Kenya | 0,28 | 0,06 | 0,89 | 0,95 |
| Germany | Sweden | 0,27 | 0,03 | 0,79 | 0,83 |
| Spain | Kenya | 0,27 | 0,36 | 0,59 | 0,95 |
| Switzerland | Zambia | 0,27 | 0,04 | 0,96 | 1,00 |
| China | El Salvador | 0,26 | 0,05 | 0,72 | 0,77 |
| India | Thailand | 0,25 | 0,10 | 0,74 | 0,84 |
| Pakistan | Philippines | 0,24 | 0,01 | 0,99 | 0,99 |
| Sweden | Kenya | 0,23 | 0,12 | 0,83 | 0,95 |
| Hungary | Ethiopia | 0,23 | 0,02 | 0,89 | 0,90 |
| China | Ethiopia | 0,23 | 0,18 | 0,72 | 0,90 |
| Zambia | Madagascar | 0,22 | 0,00 | 1,00 | 1,00 |
| Sudan | Senegal | 0,21 | 0,04 | 0,64 | 0,69 |
| Sudan | Ethiopia | 0,20 | 0,26 | 0,64 | 0,90 |
| Japan | Switzerland | 0,17 | 0,02 | 0,94 | 0,96 |
| Japan | Zambia | 0,15 | 0,06 | 0,94 | 1,00 |
| Switzerland | Pakistan | 0,13 | 0,03 | 0,96 | 0,99 |

Next, we reveal the best benchmark and learning partners network based on high export similarities, but significant differences in their REPR values (see Figure 5). While the previous network mainly shows which countries have the highest levels of export similarities, here we identify the two best benchmark countries for each country that have a high level of export similarity as well as significantly better relative ecological pollution values. To do so, we first identify the differential in the sustainability indicators from a focal country *c* in relation to all the remaining countries. A positive (negative) differential means that country *c1* has a lower (greater) sustainability indicator than a particular partner country *c2* and thus it can acquire (transfer) better practices from it. We will consider only relationships with a positive difference to draw a network with the two best partners for each country. Finally, the network is generated by taking for each focal country the two outlinks that represent the highest gain in REPR and with countries with the highest export portfolio correlation, which needs to be greater than zero.

The resulting network shows the two best benchmark and learning partners for each country. As expected, in many cases best benchmark and learning partners can be found in spatial proximity, such as Serbia learning from Bosnia and Herzegovina and Croatia, or Senegal from Zambia and Kenia, Bolivia from Panama and Peru, Kazakhstan from Russia and Ukraine, etc. However, there are also several cases in which country from one continent can also learn from countries from other continents that are able to produce similar goods, but show substantially higher levels of ecological efficiency, such as the USA learning from Japan and Singapore, or Morocco from Tunisia and Sri Lanka.

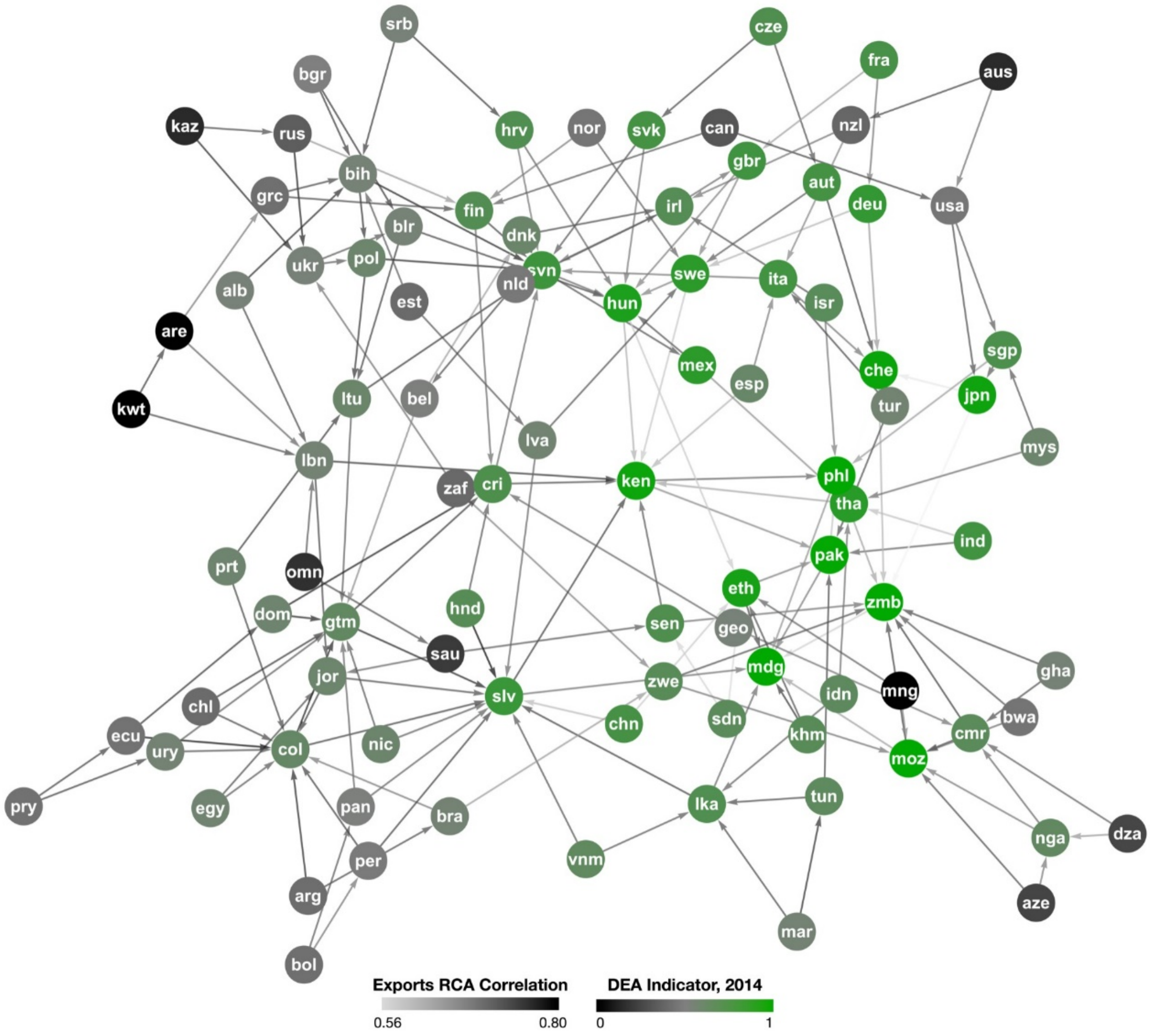

**Figure 5 –** Optimal Benchmark and Partnership Network for Sustainability Improvements.

Among the country pairs with the highest possible ecological efficiency improvements potential are, for instance, United Arab Emirates learning from Lebanon and Singapore, or Mongolia learning from Ethiopia and Sudan, the United States from Japan, or Morocco, Nicaragua and Tunisia learning from Madagascar (see Table 4). This means that countries can move beyond orienting their efficiency improvements solely based on the leading country or technology, but also have the possibility to learn from countries with similar productive structures, but significantly lower environmental damage values. This can make a difference because countries typically cannot randomly move and adopt into completely new sectors and technologies, but tend to move into activities that are similar to their previous productive portfolio (Hidalgo et al., 2007; Pinheiro et al., 2018; Hidalgo 2021). Moreover, it expands to potential learning partnerships between countries that may face similar productive challenges. This does not mean that learning from the global technology frontier and best country and technology should not also be promoted. But it provides a new layer of learning opportunities from countries with similar comparative advantages and production challenges, but that have found more efficient and ecological solutions. Our study allows for a more detailed understanding of the causes of this observation. Similar production structures indicate the existence of higher developed absorptive capacities that allow for more efficient knowledge flows and an easier exploitation of external knowledge. This not necessarily is a knowledge flow from the leading economy to the catching-up economies (Verspagen, 1992), but can be targeted on a technological level in cases of overlapping production structures.

**Table 4 –** Top 20 benchmark country pairs with highest REPR ecological efficiency improvement potential (REPR differential). The complete table can be found in Appendix A.

| Focal Country | Partner Country | Correlation | REPR Differential |
|---|---|---|---|
| Mongolia | Mozambique | 0,36 | 0,99 |
| Mongolia | Ethiopia | 0,44 | 0,90 |
| Azerbaijan | Mozambique | 0,47 | 0,72 |
| Kuwait | Lebanon | 0,49 | 0,54 |
| Botswana | Zambia | 0,46 | 0,54 |
| United Arab Emirates | Lebanon | 0,43 | 0,54 |
| Botswana | Mozambique | 0,42 | 0,54 |
| United States | Japan | 0,47 | 0,47 |
| Ghana | Zambia | 0,49 | 0,46 |
| Saudi Arabia | Senegal | 0,46 | 0,45 |
| Turkey | Pakistan | 0,47 | 0,44 |
| United Arab Emirates | Greece | 0,39 | 0,43 |
| Lebanon | Kenya | 0,54 | 0,41 |
| Nigeria | Mozambique | 0,40 | 0,38 |
| Tunisia | Pakistan | 0,48 | 0,36 |
| Norway | Sweden | 0,43 | 0,36 |
| Kazakhstan | Ukraine | 0,55 | 0,36 |
| Algeria | Cameroon | 0,44 | 0,36 |
| Spain | Kenya | 0,27 | 0,36 |
| Cameroon | Zambia | 0,50 | 0,36 |

Finally, we calculate the average relative efficiency improvement if each country would have similar efficiency values as its respective best benchmark country. The results show that countries could improve in average 22.4% of their relative efficiency if they would produce the same reduced amount of carbon dioxide and ecological footprint for a similar export portfolio than the best benchmark country. Naturally many factors, such as geography and climate conditions, institutions, closeness to supplier, and demand structures, influence the resources, energy needs and production efficiency of countries (variables that are not been considered). Nonetheless, this estimate illustrates a major potential for efficiency improvements, especially considering that despite differences in production technologies, many products (such as oranges, steel, or cars) do require similar inputs and productive capabilities across the world. So, while there are significant differences in the precise factor combination on how to produce certain products, there are also significant similarities and related efficiency differentials that can be used to identify opportunities for mutual learning and efficiency improvements.

# 5. Conclusions and policy implications

In this article, we discussed to which extent countries with similar productive structures show similarities and differences in terms of their ecological production efficiency. This analysis is important, because comparing efficiency levels of countries with very different productive specializations can cause confusion about which countries have a relatively clean production system and which countries could best learn from each other. Moreover, mere focus on aggregate indicators can also lead to political gridlock in international climate and pollution summits, where developing economies argue for the need for industrialization and thus increasing pollution levels, while some richer economies highlight their relative clean production. Comparisons based on aggregate production or pollution levels alone may not be the best way to understand which countries could best learn from each other in terms of best practices, technologies, and regulations in their industries. For instance, a car industry, a copper mine, a soybean industry, a finance industry, or a textile industry require different types of technologies and policies to move closer to the global benchmark in terms of production efficiency. Moreover, the impact of these industries also depends on the network of related industries that are present in a country. Thus, different production portfolios of countries need to be considered.

In this paper, we created a relative ecological pollution ranking (REPR) and reveal a best efficiency benchmark partner network that considers both high levels of export similarities and differences in ecological efficiency. For instance, it is not obvious from traditional efficiency rankings that the USA can learn from Japan, Cameroon from Zambia, or the United Arab Emirates from Greece. The article showed that methods from data envelopment analysis and economic complexity can identify possibilities for efficiency improvements and mutual learning better than ecological efficiency rankings based on aggregate indicators, because they consider the productive structure of each country. While having its limitations, it is a step forward in being able to compare like with like. This can also help to expand the information base and learning activities between countries with similar productive structures for the sake of a higher level of ecological efficiency. Moreover, our results indicate a major possibility of efficiency improvements within the current global production system.

The methods and insights presented here have several policy implications. First, our insights could contribute to a greater objectivity about global climate change mitigation activities. Comparing like with like significantly improves the basic conditions in international negotiations and facilitates a less distorted discussion. Second, the insights on best benchmark countries may provide valuable information for the development of international research and technology programs. It must be noted, though that in this regard our study can make a first step, but additional in-depth studies of the best benchmark countries might be necessary. For example, the information on the benchmark countries could be used to identify whether specific infrastructures or regulations are required to improve the ecological efficiency or reorganize the concerned industries. The same holds for international development programs, which might become more accurate and effective by considering REPR differentials in their policy designs. Moreover, it could be used in international investment decisions that consider environmental considerations and to merit relative levels of ecological efficiency of the potential host countries. Countries may promote investments of multinational companies (FDI) from benchmark countries with higher levels of ecological pollution to promote knowledge spillover and increase ecological efficiency. Or inputs (with similar qualities and prices) could be preferentially bought from countries with higher REPR values and/or higher sustainability standards in the supplier industries (e.g. natural resources) being enforced by large buyers (consortia).

Of course, several limitations need to be kept in mind. First, while widely used in research on productive structures, data on exports are only a proxy for productive structures of countries. They do not include non-tradables, services, internal demand, and supply structures that can significantly contribute to the overall economic output and ecological efficiency levels of countries. Nonetheless, detailed and comparable production data is not yet available for a large set of countries, and export data continues to be a valuable source of information to distinguish different national productive specializations. Moreover, it must be noted that due to converging global consumption structures, import portfolios as well as service sector portfolios tend to have lower levels of variance across countries than export portfolios. Thus, export portfolios continue to be widely available and a reliable source of information on the national productive specialization due to custom checks of both export and import countries. Moreover, the export portfolios of countries tend to indirectly depict the set of basic input factors, such as land, technology, and institutions that are necessary to be able to produce and export these goods in a competitive manner (Hidalgo and Hausmann, 2009; Hausmann et al., 2014). For instance, the export of soybeans demands a certain type of climate, while the export of robots a certain level of technological capabilities.

Another limitation is that we perform in this article a rather static framework that does not consider significant changes in terms of product diversification and the rise of new industries. Future research may need to combine both considerations of efficiency as well as likely changes in the productive portfolios of countries. Indeed, several advances have been made recently on the association between economic diversification, complexity, and sustainability (Ferraz et al., 2021). It must be noted, though, that the recent focus on green diversification opportunities should also not omit the potential efficiency improvement within the current productive specializations of countries. We show here that significant efficiency improvements would be possible within the current global production system.

There are also many political, social, and institutional issues involved that can promote or hamper the collaboration between countries that need to be considered and explored in subsequent works. For instance, many neighboring countries or best benchmark countries had political conflicts that can negatively affect knowledge transfer and mutual learning. At the same time, a common history, institutions, and language, as seen in the case of the Commonwealth countries, can help to promote communication, joint projects, and knowledge transfer. Finally, geographic factors, such as a closer or greater distance, or differences in climate conditions, can also affect the ability of countries to learn from the production systems of each other. All these considerations suggest promising paths for future research on the micro-level of cooperation between the best benchmark countries.

Despite its limitations, this article provides a new analysis framework to identify the best ecological production efficiency and benchmark countries. It can help in developing more adequate comparisons of the ecological production efficiency of countries, considering their significant differences in productive specialization, instead of merely focusing on aggregate pollution and/or GDP levels. And thus, it can help to identify which countries can best learn from each other for the sake of a cleaner global production.

# Appendix A

**Table A1**. Absolute and relative indicators for economic complexity and sustainability

| Country | Variables | | | REPR | |
|---|---|---|---|---|---|
| | ECI | $CO2_{pc}$ | EFConsPerCap | Index | Rank |
| Albania | -0.54 | 1.98 | 2.04 | 0.5263 | 69 |
| Algeria | -1.77 | 3.74 | 2.49 | 0.2786 | 90 |
| Argentina | -0.50 | 4.78 | 3.75 | 0.4313 | 83 |
| Australia | -0.85 | 15.39 | 6.75 | 0.1699 | 94 |
| Austria | 1.65 | 6.87 | 6.02 | 0.7258 | 22 |
| Azerbaijan | -1.78 | 3.93 | 2.16 | 0.2709 | 91 |
| Belarus | 0.73 | 6.70 | 4.78 | 0.5366 | 63 |
| Belgium | 0.91 | 8.33 | 6.91 | 0.5099 | 70 |
| Bolivia | -1.18 | 1.91 | 3.12 | 0.4377 | 81 |
| Bosnia and Herzegovina | 0.58 | 6.38 | 3.32 | 0.5362 | 64 |
| Botswana | -0.79 | 3.37 | 2.58 | 0.4605 | 78 |
| Brazil | -0.15 | 2.61 | 3.10 | 0.5437 | 59 |
| Bulgaria | 0.29 | 5.87 | 3.31 | 0.5035 | 73 |
| Cambodia | -0.72 | 0.44 | 1.33 | 0.6464 | 37 |
| Cameroon | -0.84 | 0.31 | 1.30 | 0.6394 | 40 |
| Canada | 0.41 | 15.16 | 7.77 | 0.3407 | 89 |
| Chile | -0.53 | 4.65 | 4.00 | 0.4246 | 84 |
| China | 1.16 | 7.54 | 3.69 | 0.7199 | 25 |
| Colombia | -0.19 | 1.79 | 2.00 | 0.6018 | 46 |
| Congo, Dem. Rep. | -0.73 | 0.06 | 1.11 | 1.0000 | 1 |
| Costa Rica | 0.09 | 1.62 | 2.52 | 0.6895 | 28 |
| Croatia | 0.84 | 3.97 | 3.62 | 0.6782 | 32 |
| Czech Republic | 1.52 | 9.17 | 5.60 | 0.6841 | 30 |
| Denmark | 0.95 | 5.94 | 7.06 | 0.5764 | 49 |
| Dominican Republic | -0.41 | 2.12 | 1.64 | 0.5636 | 54 |
| Ecuador | -1.31 | 2.75 | 2.05 | 0.4497 | 79 |
| Egypt, Arab Rep. | -0.34 | 2.23 | 1.96 | 0.5527 | 56 |
| El Salvador | -0.07 | 1.00 | 1.96 | 0.7734 | 17 |
| Estonia | 0.75 | 14.85 | 6.80 | 0.4191 | 85 |
| Ethiopia | -1.56 | 0.12 | 1.06 | 0.9045 | 11 |
| Finland | 1.50 | 8.66 | 6.03 | 0.6653 | 33 |
| France | 1.16 | 4.57 | 4.75 | 0.7156 | 26 |
| Georgia | -0.46 | 2.42 | 1.94 | 0.5309 | 66 |
| Germany | 1.81 | 8.89 | 5.03 | 0.7942 | 16 |
| Ghana | -1.47 | 0.53 | 1.90 | 0.5345 | 65 |
| Greece | -0.17 | 6.18 | 4.25 | 0.4377 | 82 |
| Guatemala | -0.41 | 1.15 | 1.79 | 0.6107 | 45 |
| Guinea-Bissau | -2.18 | 0.16 | 1.48 | 0.1633 | 96 |
| Honduras | -0.37 | 1.06 | 1.45 | 0.6532 | 35 |
| Hungary | 1.38 | 4.27 | 3.61 | 0.8873 | 12 |
| India | -0.01 | 1.73 | 1.17 | 0.7395 | 20 |
| Indonesia | -0.10 | 1.82 | 1.68 | 0.6270 | 41 |
| Ireland | 1.22 | 7.31 | 5.01 | 0.6603 | 34 |
| Israel | 1.14 | 7.86 | 4.70 | 0.6451 | 38 |
| Italy | 1.24 | 5.27 | 4.39 | 0.7368 | 21 |
| Jamaica | -0.79 | 2.58 | 1.69 | 0.5054 | 72 |
| Japan | 2.32 | 9.54 | 4.71 | 0.9391 | 9 |
| Jordan | -0.01 | 2.97 | 1.88 | 0.5703 | 52 |
| Kazakhstan | -1.01 | 14.36 | 5.72 | 0.1659 | 95 |
| Kenya | -0.52 | 0.31 | 1.05 | 0.9489 | 8 |
| Kuwait | -0.84 | 25.85 | 7.82 | 0.0000 | 99 |
| Latvia | 0.43 | 3.50 | 5.84 | 0.5491 | 57 |
| Lebanon | 0.18 | 3.84 | 3.57 | 0.5396 | 62 |
| Lithuania | 0.64 | 4.38 | 5.55 | 0.5769 | 48 |
| Madagascar | -0.82 | 0.13 | 0.97 | 1.0000 | 2 |
| Malaysia | 0.83 | 8.13 | 4.23 | 0.5684 | 53 |
| Mexico | 0.95 | 3.99 | 2.58 | 0.8293 | 14 |
| Mongolia | -1.56 | 7.09 | 7.47 | 0.0025 | 98 |
| Morocco | -0.56 | 1.75 | 1.82 | 0.5425 | 60 |
| Mozambique | -1.21 | 0.32 | 0.84 | 0.9956 | 4 |
| Netherlands | 0.76 | 9.92 | 6.13 | 0.4749 | 75 |
| New Zealand | -0.12 | 7.69 | 5.26 | 0.4056 | 87 |
| Nicaragua | -1.00 | 0.79 | 1.41 | 0.5706 | 51 |
| Nigeria | -1.72 | 0.55 | 1.17 | 0.6144 | 44 |
| Norway | 0.67 | 9.27 | 6.10 | 0.4639 | 77 |
| Oman | -0.77 | 15.19 | 6.74 | 0.1906 | 93 |
| Pakistan | -0.87 | 0.85 | 0.83 | 0.9862 | 6 |
| Panama | -0.56 | 2.26 | 2.35 | 0.5058 | 71 |
| Paraguay | -1.10 | 0.86 | 3.24 | 0.4463 | 80 |
| Peru | -0.96 | 2.05 | 2.27 | 0.4819 | 74 |
| Philippines | 0.48 | 1.05 | 1.10 | 0.9913 | 5 |
| Poland | 0.84 | 7.52 | 4.38 | 0.5732 | 50 |
| Portugal | 0.49 | 4.33 | 3.72 | 0.5627 | 55 |
| Romania | 0.79 | 3.52 | 2.78 | 0.7229 | 24 |
| Russian Federation | 0.01 | 11.86 | 5.45 | 0.3682 | 88 |
| Saudi Arabia | -0.37 | 19.44 | 6.00 | 0.2311 | 92 |
| Senegal | -0.72 | 0.62 | 1.12 | 0.6852 | 29 |
| Serbia | 0.37 | 5.28 | 2.91 | 0.5287 | 67 |
| Singapore | 1.71 | 10.31 | 5.96 | 0.6961 | 27 |
| Slovak Republic | 1.20 | 5.66 | 4.29 | 0.7247 | 23 |
| Slovenia | 1.41 | 6.21 | 4.65 | 0.7550 | 18 |
| South Africa | -0.20 | 8.98 | 3.53 | 0.4122 | 86 |
| Spain | 0.70 | 5.03 | 3.77 | 0.5898 | 47 |
| Sri Lanka | -0.37 | 0.89 | 1.49 | 0.6784 | 31 |
| Sudan | -1.84 | 0.30 | 1.27 | 0.6411 | 39 |
| Sweden | 1.65 | 4.48 | 6.50 | 0.8272 | 15 |
| Switzerland | 1.99 | 4.31 | 4.87 | 0.9567 | 7 |
| Thailand | 0.96 | 4.62 | 2.43 | 0.8384 | 13 |
| Togo | -0.46 | 0.37 | 1.07 | 0.9252 | 10 |
| Tunisia | 0.21 | 2.61 | 2.19 | 0.6219 | 42 |
| Turkey | 0.38 | 4.48 | 3.26 | 0.5475 | 58 |
| Ukraine | 0.27 | 5.02 | 2.75 | 0.5270 | 68 |
| United Arab Emirates | -0.36 | 22.94 | 10.23 | 0.0029 | 97 |
| United Kingdom | 1.40 | 6.50 | 4.71 | 0.7433 | 19 |
| United States (USA) | 1.30 | 16.50 | 8.33 | 0.4648 | 76 |
| Uruguay | -0.35 | 1.98 | 2.66 | 0.5406 | 61 |
| Vietnam | -0.13 | 1.82 | 1.79 | 0.6179 | 43 |
| Zambia | -0.54 | 0.29 | 0.97 | 0.9979 | 3 |
| Zimbabwe | -0.84 | 0.88 | 1.08 | 0.6486 | 36 |

Note: ECI = Economic Complexity Index; CO2pc = Carbon Dioxide ($CO_2$) emissions (metric tons per capita); EFConsPerCap = Ecological Footprint of consumption in global hectares (gha) divided by population.